\documentclass{arxiv}
\usepackage{lineno}

\received{xxxx, 2023}
\revised{xxxx, 2023}
\accepted{xxxx, 2024}
\submitjournal{ApJ}

\shorttitle{VLBA Observations of Parsec-scale Radio Emission in Dual AGN}
\shortauthors{Xu, Cui* et al.}
\graphicspath{{./}{figures/}}
\usepackage{amsmath}
\usepackage{hyperref}

\begin{document}
\title{Very Long Baseline Array Observations of Parsec-scale Radio Emission in Dual Active Galactic Nuclei}

\correspondingauthor{Lang Cui}
\email{cuilang@xao.ac.cn}

\author{Wancheng Xu}
\affiliation{Xinjiang Astronomical Observatory, CAS, 150 Science-1 Street, Urumqi 830011, China}
\affiliation{College of Astronomy and Space Science, University of Chinese Academy of Sciences, No.1 Yanqihu East Road, Beijing 101408, China}

\author[0000-0003-0721-5509]{Lang Cui}
\affiliation{Xinjiang Astronomical Observatory, CAS, 150 Science-1 Street, Urumqi 830011, China}
\affiliation{Key Laboratory of Radio Astronomy, CAS, 150 Science 1-Street, Urumqi 830011, China}
\affiliation{Xinjiang Key Laboratory of Radio Astrophysics, 150 Science 1-Street, Urumqi 830011, China}

\author[0000-0001-9815-2579]{Xiang Liu}
\affiliation{Xinjiang Astronomical Observatory, CAS, 150 Science-1 Street, Urumqi 830011, China}
\affiliation{Key Laboratory of Radio Astronomy, CAS, 150 Science 1-Street, Urumqi 830011, China}
\affiliation{Xinjiang Key Laboratory of Radio Astrophysics, 150 Science 1-Street, Urumqi 830011, China}

\author[0000-0003-4341-0029]{Tao An}
\affiliation{Shanghai Astronomical Observatory, CAS, 80 Nandan Road, Shanghai 200030, China}

\author{Hongmin Cao}
\affiliation{School of Electronic and Electrical Engineering, Shangqiu Normal University, 298 Wenhua Road, Shangqiu, Henan 476000, China}

\author{Pengfei Jiang}
\affiliation{Xinjiang Astronomical Observatory, CAS, 150 Science-1 Street, Urumqi 830011, China}
\affiliation{College of Astronomy and Space Science, University of Chinese Academy of Sciences, No.1 Yanqihu East Road, Beijing 101408, China}

\author[0000-0001-6947-5846]{Luis C. Ho}
\affiliation{Kavli Institute for Astronomy and Astrophysics, Peking University, Beijing 100871, China}
\affiliation{Department of Astronomy, School of Physics, Peking University, Beijing 100871, China}

\author[0000-0002-8684-7303]{Ning Chang}
\affiliation{Xinjiang Astronomical Observatory, CAS, 150 Science-1 Street, Urumqi 830011, China}

\author[0000-0002-4439-5580]{Xiaolong Yang}
\affiliation{Shanghai Astronomical Observatory, CAS, 80 Nandan Road, Shanghai 200030, China}

\author{Yuling Shen}
\affiliation{Xinjiang Astronomical Observatory, CAS, 150 Science-1 Street, Urumqi 830011, China}

\author{Guiping Tan}
\affiliation{Xinjiang Astronomical Observatory, CAS, 150 Science-1 Street, Urumqi 830011, China}

\author{Zhenhua Han}
\affiliation{College of Physics and Electronic Engineering, Xinjiang Normal University, 102 Xinyi Road, Urumqi 830054, China}

\author{Junhui Fan}
\affiliation{Center for Astrophysics, Guangzhou University, Guangzhou 510006, China}

\author{Ming Zhang}
\affiliation{Xinjiang Astronomical Observatory, CAS, 150 Science-1 Street, Urumqi 830011, China}
\affiliation{Key Laboratory of Radio Astronomy, CAS, 150 Science 1-Street, Urumqi 830011, China}
\affiliation{Xinjiang Key Laboratory of Radio Astrophysics, 150 Science 1-Street, Urumqi 830011, China}

\begin{abstract}

It is believed that dual active galactic nuclei (dual AGN) will form during galaxies merge. 
Studying dual-AGN emission can provide valuable insights into galaxy merging and evolution. 
To investigate parsec-scale radio emission properties, we observed eight radio components of four selected dual-AGN systems using the Very Long Baseline Array (VLBA) at 5 GHz in multiple-phase-center mode. 
Among them, two compact radio components, labeled J0051+0020B and J2300–0005A, were detected clearly on parsec scales for the first time. 
However, the radio emission of the other six components was resolved out in the high-resolution images. 
We provided the values or upper limits of the brightness temperature and radio emission power, and analyzed the emission origins in detail for each target. 
Based on their physical properties reported in this work and in the literature, we suggest the radio emission in J0051+0020B and J2300$-$0005A originates primarily from compact jets, while the other six sources show more complex emission mechanisms. 
In addition, our VLBA observations suggest the systematic X-ray deficit in our dual-AGN sample is likely attributed to the tidally induced effect and possible viewing angle effect.


\end{abstract}

\section{Introduction}\label{sec:intro}

The prevailing Lambda Cold Dark Matter ($\Lambda$CDM) cosmological model describes a hierarchical structure formation process in the Universe. 
Small galaxies assemble through mergers and interactions into progressively larger systems, eventually forming the massive galaxies and clusters we observe today \citep{2005Simulating}. 
It is generally believed that each galaxy typically harbors a supermassive black hole (SMBH) at its center \citep{kormendy2013coevolution}. 
During galaxy mergers, tidal torques could trigger the accretion and feedback of central black holes \citep{1996model1, 2005DiMatteo}, but rare dual active galactic nuclei (dual AGN) could be observed \citep{Van2012}. 
These dual AGN serve as essential targets for multiwavelength astronomical observations and offer crucial insights into the processes of black hole mergers and activations.
For example, some observations suggest that galaxy mergers could trigger black hole accretion \citep[e.g.,][]{liuxin2012ApJ, Koss2012, 2018fu}. However, recent surveys suggest the stochastic accretion mode could play a dominant role even in the presence of kiloparsec-scale dual AGN systems \citep[e.g.,][]{hopkins2006fueling, 2019ApJ...883...50G, Steffen2023ApJ}.
Moreover, galaxy mergers can trigger star formation, redistribute gas and dust, and alter galaxy morphology, making dual AGN a unique observational window into galaxy mergers and evolution \citep[e.g.,][]{liuxin2012ApJ}. 
Dual AGN exhibit detectable signatures across the electromagnetic spectrum due to their unique radiative properties. 
These features, observable with current capabilities, make dual AGN compelling targets for investigating galaxy evolution and have inspired diverse research programs to unravel the intricate dynamics and linkage between galaxies and their central SMBHs (e.g., \citealt{komossa2002discovery}; \citealt{bianchi2008chandra}; \citealt{green2010sdss}; \citealt{fu2015binary}; \citealt{2017MNRAS.471.1873Y}; \citealt{chen2023close}; \citealt{gross2023testing}).

However, observational and interpretative challenges are significant due to the complexity of dual AGN. 
Anomalous jet morphologies, such as X-shaped radio galaxies  \citep[e.g.,][]{merritt2002tracing}, double-double radio galaxies \citep[e.g.,][]{schoenmakers2000radio}, and ring structure \citep[e.g.,][]{roos1994massive} or spiral trajectory \citep[e.g.,][]{Britzen_2012} in radio jets, can indicate dual AGN. These structures can arise from the intense tidal torques during galaxy mergers, which alter the jet's collimated structure. 
Other studies search for double-peaked emission lines from clouds in broad or narrow-line regions associated with the SMBH binary's orbital motion \citep[e.g.,][]{smith2012double}. 
However, almost all of the double-peaked AGNs are caused by gas kinematics rather than true dual-AGN systems \citep[e.g.,][]{bogdanovic2009emission, fu2012ApJ}.
For closely separated binaries, cyclic outburst activities \citep{lehto1996oj} provide alternative indicators, although they also have limitations. Given the complex nature of dual AGN, reliable confirmation requires multiple observational clues, as no single signature guarantees definitive evidence.
A multiwavelength and multifaceted approach, leveraging observations across the electromagnetic spectrum from radio to X-rays, and analyzing the jet morphology, spectral line profiles, and time-domain behavior, is required to disentangle the intricate workings of these systems.

A useful strategy involves assembling dual AGN candidates based on spatially resolved optical double-peaked profiles and validating these candidates through high-resolution imaging.
Recent Very Large Array (VLA) Stripe 82 survey and optical spectroscopy have revealed numerous potential dual AGN, and their binary structures have been verified by follow-up high-resolution VLA observations \citep{fu2015radio,fu2015binary,gross2023testing}. 
Very Long Baseline Interferometry (VLBI) provides the unique angular resolution necessary to characterize their radio properties on parsec scales and is crucial in studying the origins of radio emission.

In this research, we used the Very Long Baseline Array (VLBA) to conduct high-resolution imaging observations of the four dual AGN identified by \citet{fu2015binary}, which enabled us to probe the innermost structures of these dual AGN systems, ascertain the presence of compact radio cores, and examine the morphology and spatially resolve extended features like jets.
Section \ref{target selection} describes our target selection and some basic properties of dual-AGN samples. 
Section \ref{Observation} introduces the specific VLBA observations. 
The details of the VLBI data reduction are given in Section \ref{Data Redunction}. 
We present our results and the discussion in Section \ref{Results} and Section \ref{Discussion}, respectively.
Finally, we summarize our findings in Section \ref{Conclusions}. 
A flat $\Lambda$CDM cosmological model with $H_0$ = 67.7 km s$^{-1}$ Mpc$^{-1}$, $\Omega_m$ = 0.31, and $\Omega_\Lambda$ = 0.69 is adopted throughout this paper \citep{2020A&A...641A...6P}.

\section{target selection} \label{target selection}

Sloan Digital Sky Survey (SDSS) Stripe 82 covers a continuous area of approximately 300 square degrees, centered on the celestial equator with right ascension (RA) between 20 and 4 hours and declination (Dec) between $-1.26$ and $+1.26$ degrees. 
This area has been repeatedly scanned 70--90 times by the SDSS imaging survey, resulting in data that are $\sim$2 mag deeper than the best single-epoch SDSS data \citep{jiang2014sloan}. 
The significant improvement in data depth and image quality makes Stripe 82 ideal for searching dual AGN, which is necessary to identify the optical counterparts of these systems.
In addition, about one-third of the Stripe 82 area has been mapped by the VLA at 1.4 GHz mostly in the A-configuration \citep{hodge2011high}. 
The VLA Stripe 82 radio survey has an angular resolution of 1.8 arcsec and a median root mean square (RMS) noise of 52 $\mu$Jy/beam, significantly surpassing the five-arcsecs resolution and 150 $\mu$Jy/beam median RMS noise of the Faint Images of the Radio Sky at Twenty-centimeters (FIRST) Survey in the VLA B-configuration \citep{becker1995first}. 
The superior data quality and resolution of the VLA Stripe 82 survey make it well-suited to search for dual-core AGN in the radio images directly. 
Utilizing these advantages, \citet{fu2015radio} selected the Stripe 82 sky region to perform a systematic search for dual AGN. 
In addition, Stripe 82 has extensive multi-wavelength coverage, enabling a multi-band study of the detected dual-AGN candidates \citep[e.g.,][]{Martin_2005,arnaboldi2007eso,lawrence2007ukirt,hodge2011high}. 

By comparing the VLA and SDSS images, \citet{fu2015radio} identified 52 dual-AGN candidates, with six confirmed through optical spectroscopy. 
The six dual-AGN candidates exhibit spatially resolved compact binary structures in optical and radio bands, with angular separations between 1.6 arcsec and 4.9 arcsec (or 4--12 kiloparsecs). 
All 12 galaxies in these six systems show evidence of AGN activity based on either excess radio power relative to H$\alpha$ luminosity or optical emission line ratios. While their radio morphologies are generally compact, some show signs of extended structures. 
A combination of VLA, SDSS, and Keck Low-Resolution Imaging Spectrometer (Keck/LRIS) observations reveal a mixture of AGN--star formation in the optical diagnostic diagrams, as opposed to pure star formation.

\citet{fu2015binary} presented follow-up VLA C-band observations in the A-configuration with 0.3-arcsec resolution of the six dual-AGN candidates discovered by \citet{fu2015radio}. 
Two pairs are determined to be line-of-sight projections, while the higher-resolution radio data confirmed the remaining four as true dual AGN based on their compact, steep-spectrum radio emission consistent with synchrotron radiation from sub-kiloparsec scales. 
In the follow-up research, \citet{2019ApJ...883...50G} reveal the accretion properties of the four confirmed dual AGN through Chandra X-ray observations and provide their more precise radio and spectroscopic properties \citep{gross2023testing}.
The detailed information of four confirmed dual AGN in VLA C-band observations is shown in Table \hyperlink{tab1}{1}. 
However, VLBI observations are still needed to characterize the compactness and origins of their radio emission on milliarcsec (mas) scales near the galactic nuclear regions. 
We aim to investigate the radio emission properties of the four confirmed dual-AGN samples by VLBI, such as angular structures, flux densities, brightness temperatures, and emission powers, and to identify further their radio emission origins, contributing to our understanding of galaxy evolution and merging.

\hypertarget{tab1}{}
\begin{deluxetable*}{ccccccc}[h]
\tablenum{1}
\tablecaption{Four dual AGN in VLA C-band observations}
\tablewidth{0pt}
\tablehead{
\colhead{Target} & \colhead{Component} & \colhead{Radio Designation} &
\colhead{Redshift} & \colhead{Separation} & \colhead{${S_{\rm5GHz}^{\mathrm{VLA}}}$} & \colhead{${\alpha_{\rm{4-8GHz}}}$}\\
\colhead{} & \colhead{} & \colhead{(J2000)} & \colhead{} &
\colhead{(kpc)} & \colhead{(mJy)} & \colhead{} \\
\colhead{(1)} & \colhead{(2)} & \colhead{(3)} &
\colhead{(4)} & \colhead{(5)} & \colhead{(6)} & \colhead{(7)}
}
\startdata
J0051+0020& A &005114.11 ${+}$002049.5 &0.11257&7.1 & 1.107 $\pm$ 0.025 & ${-}$0.93 $\pm$ 0.09\\
& B & 005113.93 ${+}$002047.2 & 0.11253 &   7.1  & 0.360 $\pm$ 0.021 & ${-}$1.08 $\pm$ 0.22\\
J2206+0003 & A & 220635.08 ${+}$000323.2 & 0.04640 & 4.1 & 0.715 $\pm$ 0.022 & ${-}$1.14 $\pm$ 0.12\\
& B & 220634.98 ${+}$000327.6 & 0.04656 &  4.1   & 0.110 $\pm$ 0.043 & ${-}$1.69 $\pm$ 1.67\\
J2232+0012 & A & 223222.60 ${+}$001224.7 & 0.22187 & 11.6 & 0.307 $\pm$ 0.021  & ${-}$1.29 $\pm$ 0.30\\
& B & 223222.44 ${+}$001225.9 & 0.22128 &   11.6   & 0.193 $\pm$ 0.032  & ${-}$1.80 $\pm$ 0.73\\
J2300${-}$0005 & A & 230010.18 ${-}$000531.6 & 0.17971 & 7.7 & 0.563 $\pm$ 0.020 & ${-}$0.52 $\pm$ 0.14\\
& B & 230010.24 ${-}$000533.9 & 0.17981 &  7.7   & 0.084 $\pm$ 0.028 & ${-}$1.16 $\pm$ 1.44 \\
\enddata
\tablecomments{Each pair of rows represents two components of one dual AGN. In each pair, the component with higher flux density is denoted as A, while the other one is denoted as B. Column 3: J2000 radio designation of each component. Column 4: redshift. Column 5: the separation between two components of each dual AGN. Column 6: the inferred VLA 5-GHz flux density and error based on the VLA C-band observations. Column 7: the VLA spectral index and error between 4-8 GHz. The data refer to \citet{fu2015binary}.}
\end{deluxetable*}

\section{Observations} \label{Observation}

We observed dual-AGN samples using the VLBA at 5 GHz on February 8, 2018, and February 11, 2018. 
The observational data were collected at a rate of 2048 Megabits per second at each participating radio telescope, using left-handed circular polarization (LCP), 4 intermediate frequency channels (IF), and 128-MHz bandwidth per baseband. 
The data streams were correlated in the distributed FX software correlator (DiFX; \citealt{deller2007difx}) with a 1-second integration time and 256 spectral channels per IF. 
The experiment was separated into two segments with project codes BL255A and BL255B. 

The VLBA observations were conducted in phase-referencing mode \citep{beasley1995vlbi} and multiple-phase-center mode \citep[e.g.,][]{deller2011difx,Middelberg2011,Morgan2013,Deller2014}, the latter can provide high-resolution imaging of any position within the primary beam area (FWHM$\sim$10$\arcmin$ at 5 GHz). 
A similar VLBA 5-GHz observation was previously carried out for a dual AGN at the milli-Jansky (mJy) level by \citet{wrobel2014evidence}, demonstrating the feasibility of sub-mJy imaging with the VLBA in phase-referencing mode. 
The radio telescopes were pointed to the priori positions of component A, the primary radio cores of the dual AGN inferred from the VLA C-band observations (see Table \hyperlink{tab1}{1}). 
The total observing time for each target source and phase calibrator combination was about 135 minutes. The phase-referencing cycle time was 9 minutes long, with 6 minutes spent on the target source. Therefore, approximately 90 minutes were spent on each target. 
Except for the four primary radio cores of the dual AGN, the radio positions of secondary cores B for correlation are presented in Table \hyperlink{tab1}{1}. 
The suitable phase calibrators (see Table \hyperlink{tab2}{2}), within about 1-2° angular separation from the respective targets, were selected from the Astrogeo database (rfc$_-$2017c)\footnote{\url{http://astrogeo.org}}. 
The bright fringe-finder radio source used in this experiment was J2212+2355. 
The observing parameters, including the names of radio telescopes participating in the project segments, are listed in Table \hyperlink{tab2}{2}.

\hypertarget{tab2}{}
\begin{deluxetable*}{ccccccc}[h]
\tablenum{2}
\tablecaption{The VLBA 5-GHz observing parameters}
\tablewidth{0pt}
\tablehead{
\colhead{Project ID} & \colhead{Date} & \colhead{Band} &
\colhead{Target} & \colhead{Phase Calibrator} & \colhead{Fringe Finder} & \colhead{Antenna} \\
\colhead{(1)} & \colhead{(2)} & \colhead{(3)} &
\colhead{(4)} & \colhead{(5)} & \colhead{(6)} & \colhead{(7)}
}
\startdata
BL255A & 08-Feb-18 & C-band  & J0051+0020 & J0052 +0035 & J2212 +2355 & BR, FD, HN, KP, LA, MK, NL, OV, PT \\
       &           &   & J2206+0003 & J2206 $-$0031 &            &  \\
\hline
BL255B & 11-Feb-18 & C-band & J2232+0012 & J2229 +0114 & J2212 +2355 &  BR, FD, KP, LA, MK, NL, OV, PT\\
       &           &   & J2300$-$0005 & J2254 +0054 &            &  \\
\enddata
\tablecomments{Telescope codes: BR - Brewster (Washington), FD - Fort Davis (Texas), HN - Hancock (New Hampshire), KP - Kitt Peak (Arizona), LA - Los Alamos (New Mexico), MK - Mauna Kea (Hawaii), NL - North Liberty (Iowa), OV - Owens Valley (California), PT - Pie Town (New Mexico). Using LA as a reference antenna. Some antennas did not participate in the observations because of bad weather.}
\end{deluxetable*}

\section{Data Reduction} \label{Data Redunction}

The data calibration was conducted in the NRAO Astronomical Image Processing System (AIPS, \citealt{greisen2003aips}), and generally followed the guide to VLBA data calibration in AIPS COOKBOOK\footnote{\url{http://www.aips.nrao.edu/cook.html}}. 
The Earth Orientation Parameters (EOP) correction was done using the latest EOP file. 
Parallactic angle correction for the altitude-azimuth mounted antennas, digital sampling correction, and ionospheric correction were applied. 
Manual phase calibration and global fringe fitting were performed. The phase, delay, and delay-rate solutions obtained for the phase-reference calibrators were interpolated and applied to the target sources. 
The bandpass correction was performed using the corresponding fringe finder source. 
Finally, a priori amplitude calibration was performed using the antenna gain curves and the system temperatures measured at the VLBI stations during the system's observations. 
For the calibration of the B components (the weaker sources) in the systems, due to the use of multiple-phase-center technology, we copied and applied the solution (SN) tables and bandpass (BP) tables of the A components (the brighter sources) after calibration to data of secondary cores and interpolated the phase, delay, and delay-rate solutions obtained. 
We used the Los Alamos (LA) telescope throughout the calibration procedure as the reference antenna.

After calibration, the calibrated visibility data for each target source were exported from AIPS and loaded into the DIFMAP \citep{shepherd1997difmap} software package for imaging and fitting the brightness distribution models. 
In DIFMAP, we searched for possible signals within the range of $\pm$1 arcsec from the prior positions of the target sources. We detected two reliable sources (J0051+0020B and J2300$-$0005A) with signal-to-noise ratios (SNRs) exceeding 6.
Through moved phase-referencing centers to their peak-flux positions using {\tt\string CLCOR}, their SNRs were improved.
We used one circular Gaussian modeling through {\tt\string MODELFIT} to allow for a quantitative description of the source structures. 
This method enabled us to obtain estimates for component sizes, positions, and flux densities. 
We did not attempt phase and amplitude self-calibration for the detected sources, due to their weak brightness. 

\begin{figure}[ht!]
\centering
\includegraphics[height=8cm, width=17.45cm]{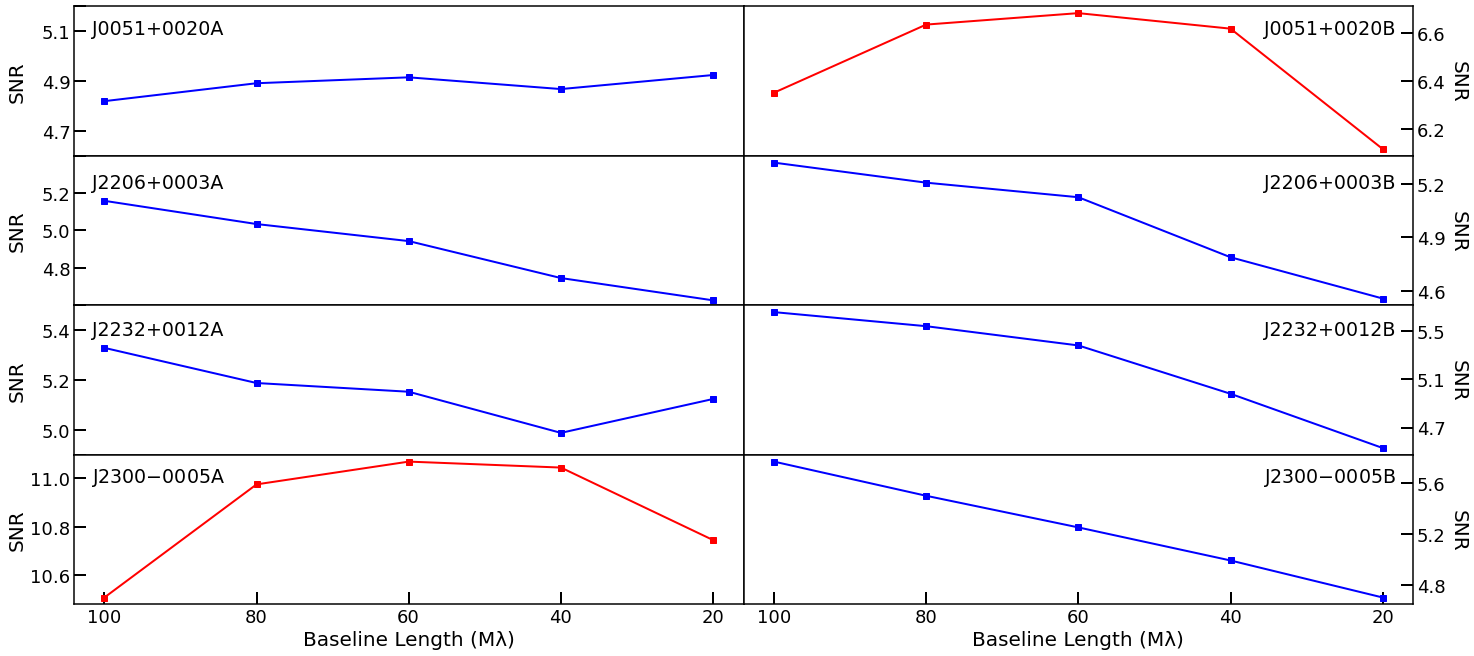}
\caption{The changes of SNRs in dirty images after reducing the weight of the long baseline. The imaging range is ±1 arcsec. The horizontal axis represents the baseline length of 50\% weight in a  Gaussian taper, and the vertical axis represents the SNR after changing the weight. Among them, the red lines represent the changes in SNR of the compact cores J0051+0020B and J2300$-$0005A, showing a clear trend of increasing first and then decreasing. The blue lines belong to undetected sources, and their SNRs do not change significantly or show a continuous downward trend. \label{fig1}}
\end{figure}

The other sources (J0051+0020A, J2206+0003A\&B, J2232+0012A\&B, and J2300$-$0005B) have an SNR of less than 6. The peak seen in the respective dirty image is likely noise rather than reliable detection.
To improve the chance of detecting extended emission, we attempted to increase the weighting on the short (u, v) spacing and reduce the resolution by applying a two-dimensional Gaussian taper to the visibility data \citep[e.g.,][]{liu2022vlba}. 
This gradually reduces the weight to 50\% at UV distance (100, 80, 60, 40, 20)$\times$M$\lambda$ (in units of wavelength). 
As the resolution decreased, the amount of data used for imaging also decreased, impacting the sensitivity of the data. 
When the weight at 20M$\lambda$ UV distance is reduced to 50\%, the noises in the dirty images have been greater than 0.3 mJy/beam. 
Therefore, the SNRs of two detected sources experienced a slight increase and then decreased, after the weight of the short (u, v) baseline was increased by tapering.
However, still no reliable signal was found at the radio positions of other sources and surrounding $\pm1$-arcsec areas.
In the process, the changes of SNRs in dirty images are shown in Figure \ref{fig1}.
The significant differences between the two VLBA-detected sources and other sources support the result of two reliable detected sources (J0051+0020B and J2300$-$0005A) and six undetected sources.

\section{Results} \label{Results}

\begin{figure}[ht!]
\centering
\includegraphics[height=19.6cm, width=15.1cm]{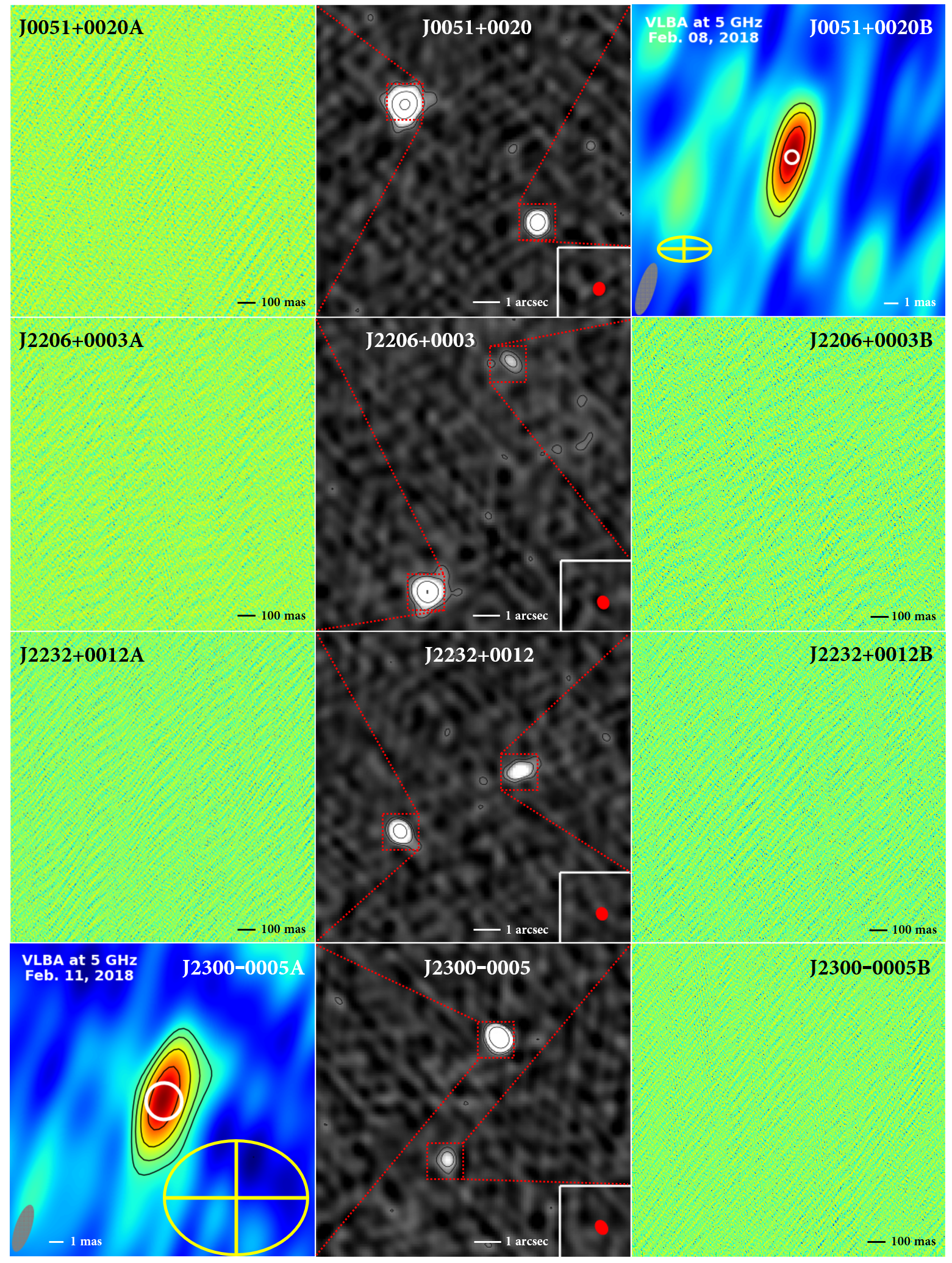}
\hypertarget{fig2}{}
\caption{Naturally weighted VLBA 5-GHz images (side columns) and VLA C-band images (middle column) of target sources. The black contours in the middle column are at (+3, +6, +24, +96) × $\sigma$, and the beam is in the bottom right corner \citep{fu2015binary}. For the compact sources J0051+0020B and J2300$-$0005A, the final Gaussian fitting images with a scale of ±10 mas are provided, and the contours are at (+1, +$\sqrt{2}$, +2, +2$\sqrt{2}$) × 3$\sigma$. The white circle in the center of the image provides the angular scale of Gaussian fitting, and the beam is in the bottom left corner. No reliable sources were found within the range of ±1 arcsec for the other extended components, so only dirty images are provided. The yellow crosses and ellipses represent the ±1$\sigma$ region around the peak positions in Gaia DR3 (\url{https://gea.esac.esa.int/archive/}). The dirty images of undetected sources are drawn centered on their VLA C-band locations.}
\end{figure}

Our VLBA 5-GHz observations detected two radio cores, and the other six components were resolved out.
Given that our observation frequency is close to the VLA C-band observations and the steep spectra of targets (see Table \hyperlink{tab1}{1}), the effect of synchronous self-absorption (SSA) on our VLBA detection results could be insignificant.
While relativistic beaming effects can amplify or diminish the apparent brightness of different jet regions, thus affecting the detectabilities in the VLBI observations \citep[e.g.,][]{beaming}.
The AGN with face-on jets could be augmented significantly by the beaming effect and easily detected.
In contrast, the beaming brightening of edge-on or reverse jets could be insignificantly and they even be Doppler dimmed.
It is noted that two VLBA-detected sources do not have the highest flux densities in the VLA observations.
The VLBA non-detection of the sources with higher VLA flux densities is probably due to their extended emission or insignificant beaming effects. 
In addition, the lack of detection of J2206+0003B and J2300$-$0005B may be due to the insufficient sensitivity of our VLBA observations. 
Because their expected 5-GHz flux densities are both lower than the VLBA 6$\sigma$ detection threshold ($\sim$0.15 mJy).

The two VLBA-detected radio cores could both have face-on jets (both have high compactness, see Section \ref{discussion2}).
Among them, J0051+0020B has both active accretion and jet activity, given its high Eddington ratio ($\sim$0.1) and clear detections in VLBA and Chandra \citep{fu2015binary,2019ApJ...883...50G}.
On the contrary, J2300$-$0005A is determined as a Low Ionization Nuclear-Emission Region (LINER) with a low Eddington ratio ($\sim$10$^{-4}$) and a high black hole mass ($\sim$10$^9 M_{\odot}$) in the BPT [N II] classification \citep{gross2023testing}.
Its ability to reach a high enough radio luminosity to be detected by the VLA and VLBA could be related to the high mass of the black hole.
The comparative properties of the two compact sources suggest that VLBA detectability (whether there is a compact radio component) does not seem biased towards a particular AGN population.

The {\tt\string MODELFIT} images of the detected sources and the dirty images of the undetected sources, as shown in Figure \hyperlink{fig2}{2}, are imaged with natural weighting. 
In Section \ref{compact}, the precise phase-referenced positions of two radio cores are provided.
The radio emission properties of two detected radio cores are examined in Section \ref{model}, including their flux densities, structures, brightness temperatures, and emission powers. 
For the other six undetected sources, we provide their upper limits of flux densities, brightness temperatures, and radio emission powers on parsec scales. 
The VLBA 5-GHz radio properties of all target sources are shown in Table \hyperlink{tab4}{4}.

\subsection{Phase-referenced Positions} \label{compact} 

Phase-referenced datasets were used to measure the astrometric positions of two detected sources (J0051+0020B and J2300$-$0005A). 
This is feasible because the coordinates of the reference sources are accurately known in the Astrogeo database.
The two VLBA-detected sources were imaged in DIFMAP, and then the AIPS program {\tt\string TVMAXFIT} was used to derive the coordinates of their brightness peaks. 
The high-resolution VLBA observations provide precise positions of radio cores (see Table \hyperlink{tab3}{3}). 
The positional uncertainties arise from a mixture of the positional uncertainties of phase calibrators, the thermal noises of the interferometer phases, and systematic errors scaled by the target-calibrator separations \citep{pradel2006astrometric}. 
We estimate the right ascension and declination coordinates provided in Table \hyperlink{tab3}{3} to be accurate within 1.5 mas for J0051+0020B and 0.5 mas for J2300$-$0005A, respectively.

\hypertarget{tab3}{}
\begin{deluxetable*}{cccc}[h]
\tablenum{3}
\tablecaption{Phase-referenced positions of two VLBA-detected sources}
\tablewidth{0pt}
\tablehead{
\colhead{Source} & \colhead{RA (J2000)} & \colhead{Dec (J2000)} & \colhead{Pos. err.} \\
\colhead{} & \colhead{(h m s)} & \colhead{(° ' ")} & \colhead{(mas)}
}
\startdata
J0051+0020B & 00:51:13.9342632 &  +00:20:47.177367  & ±1.302 \\
J2300$-$0005A & 23:00:10.1779370 & $-$00:05:31.586715 & ±0.248 \\
\enddata
\tablecomments{Column 2--3: astrometric positions (right ascension and declination) of two VLBA-detected sources measured by phase referencing. Column 4: position error (1$\sigma$).}
\end{deluxetable*}

The VLA C-band observations did not provide accurate positional information for the targets. 
Consequently, our search for the target signals was conducted within the ±1-arcsec range of the phase centers. 
For two sources detected in VLBA observations, the positions of the peak signals are within 70 mas from the phase centers. 
The detected signals have SNRs exceeding 6, and no other signals are found within the fields of view. 
This is a clear detection of two single radio cores from two different dual AGN. 
The positional deviations of the correlation centers could impact the SNRs. 
Therefore, we adjusted the correlation centers from the prior phase centers to the VLBA peak-flux positions using the {\tt\string CLCOR} command in AIPS. 
The adjustments improved the SNRs of two detected sources and further confirmed the reliability of detection.

\subsection{Images and Brightness Distribution Models\label{model}} 

The high-resolution VLBA observations identified two sources, corresponding to J0051+0020B (SNR$>$7) and J2300$-$0005A (SNR$>$10), while the other target sources are not reliably detectable with our current observational capabilities.
J0051+0020B is unresolved in the VLBA image, and its angular scale is smaller than the synthesized beam and theoretical resolution limit ($\approx$1.31 mas, \citealt{lobanov2005resolution}).
Another detected source J2300$-$0005A is slightly resolved in the VLBA image but has a low flux density (S$_\nu$$<$0.5 mJy). 
Therefore, one single circular Gaussian model was used to fit the source brightness distribution for each detected source. 
The errors of the model parameters are estimated according to \citet{lee2008global}, assuming that the errors are stochastic and independent for each parameter estimated for the component \citep{fomalont1999image}. 
For the flux densities, we assume an additional 10\% error added in quadrature to account for the uncertainties of the VLBI amplitude calibration that is based on antenna system temperatures and gain curves. 
The uncertainties of angular scales are calculated according to the synthesized beam sizes and the SNRs of the components. 
For the six undetected target sources, we estimated their lower limits of angular scales based on the synthesized beam sizes and provided their 6$\sigma$ upper limits of flux densities.
It is noted that, for J2206+0003B and J2300$-$0005B, we use the estimated VLA 5-GHz flux densities as their upper limits, due to the low sensitivities of our VLBA observations and the 6$\sigma$ upper limits being higher than the VLA flux densities.
The VLBA imaging and model parameters are listed in Table \hyperlink{tab4}{4}.

\hypertarget{tab4}{}
\begin{deluxetable*}{ccccccccc}[h]
\tablenum{4}
\tablecaption{VLBA 5-GHz radio properties of eight components in four dual AGN}
\tablewidth{0pt}
\tablehead{
\colhead{Source} & \colhead{$S^{\rm Peak}_{\rm5GHz}$} & \colhead{RMS} &
\colhead{SNR} & \colhead{Beam size} & \colhead{$S_{\rm5GHz}^\mathrm{VLBA}$} & \colhead{$\theta$} &
\colhead{log $T_b$} & \colhead{log $L_{\rm5GHz}$} \\
\colhead{} & \colhead{(mJy/beam)} & \colhead{(mJy/beam)} & \colhead{} & \colhead{(mas×mas)} & \colhead{(mJy)} & \colhead{(mas)} & \colhead{(K)} & \colhead{(W/Hz)} \\
\colhead{(1)} & \colhead{(2)} & \colhead{(3)} &
\colhead{(4)} & \colhead{(5)} & \colhead{(6)} & \colhead{(7)} &
\colhead{(8)} & \colhead{(9)}
}
\startdata
J0051+0020A & 0.121 & 0.024  & 4.99 & 7.04×2.04 & $<$0.146 & $>$3.79 & $<$5.74 & $<$21.72 \\
J0051+0020B & 0.183 & 0.023 & 7.85  & 7.04×2.04 & 0.20±0.03 & 0.84±0.24 & $\sim$7.21 & $\sim$21.77 \\
J2206+0003A & 0.121 & 0.023  & 5.16 & 6.13×1.93 &$<$0.140 & $>$3.44 & $<$5.78 & $<$20.85 \\
J2206+0003B & 0.132 & 0.024  & 5.50 & 6.13×1.93 & $<$0.110$^*$ & $>$3.44 & $<$5.67 & $<$20.75 \\
J2232+0012A & 0.139 & 0.026  & 5.33 & 6.24×2.04 &$<$0.156 & $>$3.57 & $<$5.86 & $<$22.35 \\
J2232+0012B & 0.156 & 0.026  & 5.98 & 6.24×2.04 &$<$0.157 & $>$3.57 & $<$5.87 & $<$22.35 \\
J2300$-$0005A  & 0.248 & 0.024 & 10.07 & 6.33×2.06 & 0.40±0.05 & 2.37±0.18 & $\sim$6.64 & $\sim$22.49 \\
J2300$-$0005B & 0.147 & 0.026  & 5.64 & 6.33×2.06 & $<$0.084$^*$ & $>$3.61 & $<$5.57 & $<$21.88 \\
\enddata
\tablecomments{Column 2--5: VLBA imaging parameter. For two detected sources J0051+0020B and J2300$-$0005A, the peak flux densities are extracted from {\tt\string MODELFIT} images after Gaussian model fitting, and the RMSs are extracted from residual images. For the other undetected sources, their SNRs are all extracted from dirty images. Column 6: the flux density or its 6$\sigma$ upper limit. For J2206+0003B and J2300$-$0005B marked `*', replace with their expected VLA 5-GHz flux-density upper limits. Column 7: fitted Gaussian component diameter (FWHM). For undetected sources, use $\theta = \sqrt{beam_1 \times beam_2}$ to represent their lower limits of angular scales. Column 8: rest-frame brightness temperature or the upper limit. Column 9: radio emission power (monochromatic luminosity) or the upper limit.}
\end{deluxetable*}

We estimated the brightness temperatures and radio emission powers or their upper limits in the rest frames of target sources based on the flux densities and angular scales, and also listed in Table \hyperlink{tab4}{4}.
The brightness temperatures or their upper limits of target sources are estimated using the following formula \citep{condon1982strong,lee2008global}:

\begin{equation}\label{formula1}
T_b = 1.22 \times 10^{12}(1+z) \frac{S_\nu}{\theta^2 \nu^2} \quad \mathrm{[K]}
\end{equation}

where $z$ is the redshift, $S_\nu$ is the flux density in [Jy], $\nu$ is the observing frequency in [GHz], and $\theta$ is the full width at half-maximum (FWHM) size of the circular Gaussian model component in [mas]. 
The radio emission powers (monochromatic luminosities) or their upper limits in the rest frames of target sources are estimated using the following formula \citep{hogg2002k}:

\begin{equation}\label{formula2}
L_\nu = 4\pi d_L^2 \frac{S_\nu}{(1+z)^{1+\alpha}} \quad \mathrm{[W/Hz]}
\end{equation}

where $d_L$ is the luminosity distance and $\alpha$ is the spectral index ($S_\nu$$\propto$$\nu^\alpha$). Due to the lack of VLBA-scale spectral indices of the targets and their low redshifts ($z$$<$0.2), we approximated $z$$=$0 and estimated the emission-power upper limits of the undetected sources. For the two VLBA-detected sources, we assumed a flat spectral index ($\alpha$=0) on VLBA scales \citep[e.g.,][]{wang2023vlbi2} according to their VLA spectral index maps by \cite{fu2015binary}, and estimated the approximate values of their radio emission powers (see Table \hyperlink{tab4}{4}).
The two VLBA-detected radio cores have high brightness temperatures and radio emission powers, similar to typical jet-dominated AGN ($T_b$$>$$10^6$ K, $L_\nu$$>$$10^{21}$ W/Hz; \citealt{homan2006intrinsic}).
Given their low flux densities, the jets should be weak rather than the highly relativistic ones typically observed in radio-loud AGN.
The other sources have no detectable parsec-scale radio structures.
Therefore, they could have kiloparsec-scale weak and extended jets instead of parsec-scale compact jets.

\section{Discussion} \label{Discussion}

\subsection{Astrometric Analysis} \label{discussion1}

After obtaining the precise positions of two VLBA-detected sources, we compared their phase-referenced positions (see Table \hyperlink{tab3}{3}) with their positions in Gaia Data Release 3 (Gaia DR3, \citealt{Gaia2016, Gaia2023}).
The VLBA-Gaia separation of J0051+0020B is about 9.32 mas ($\sim$19.1 parsec) and for J2300$-$0005A is about 8.19 mas ($\sim$24.9 parsec, see Figure \hyperlink{fig2}{2}). 
Compared to the positional uncertainties of the VLBA ($<$1.5 mas for J0051+0020B; $<$0.5 mas for J2300$-$0005A) and Gaia DR3 ($<$1.95 mas for J0051+0020B; $<$6.35 mas for J2300$-$0005A), the VLBA-Gaia separations are significant.
The above two VLBA-detected sources have large $astrometric_{-}excess_{-}noise$ (AEN) in the Gaia DR3 catalogue, which indicates their significant disagreements between Gaia observations and the best-fitting standard astrometric models \citep{Lindegren2021}.
The AEN has a significant dependence on Gaia magnitude \citep[e.g.,][]{wang2023varstrometry}.
Therefore, the large AENs in our target sources could be caused by their dark Gaia $g$-$band$ magnitudes ($>$20).
Moreover, the extended host emission could also have a strong effect on the accuracy of astrometric variability \citep[e.g.,][]{Hwang2020}, given the low redshifts ($z<$0.2) of our targets.

In the SDSS images, the host galaxies of some dual AGN exhibit significant distortion and association, such as J0051+0020A\&B \citep{fu2015radio,fu2015binary}.
Therefore, the Gaia position may not correspond to the actual location of the black hole or accretion disk. 
The galactic merging could also be the reason for the significant VLBA-Gaia position offsets and the forming of the off-nucleus AGN \citep[e.g.,][]{Chen2023}.
The more significant accretion and star formation of J0051+0020B compared to J2300$-$0005A suggest the stronger impact of galactic merging on J0051+0020B \citep{fu2015binary}.
The more significant VLBA-Gaia position offset of J0051+0020B (about 3$\sigma$) compared to J2300$-$0005A (about 1.5$\sigma$) also supports the inference.

\subsection{Radio Emission Origins} \label{discussion2}

The radio emission from radio-quiet AGN is often thought to be associated with less energetic jets, AGN-driven outflows, corona, and star formation within the host galaxy \citep[see the recent review by][]{panessa2019origin}. 
Different emission mechanisms dominate on different physical scales, manifesting distinct observational signatures. 
These various radio emission components can blend but can be disentangled based on morphology, size, brightness temperature, spectral index, and other observational diagnostics \citep{wang2023vlbi2}. 

Compact radio cores are typically detected for jet-dominated emission in radio-quiet AGN. 
These compact cores can exhibit high brightness temperature ($\gtrsim$10$^7$ K), and their detection in VLBI observations provides strong evidence for non-thermal emission, helpful in distinguishing between AGN-driven and star formation-driven radio emission \citep{blundell1998central,wang2023vlbi, wang2023vlbi2}. 
In the VLA C-band observations, all eight target sources have detected radio emission which can be partly associated with extended jet structures at kiloparsec scales \citep{silpa2021outflows}.
Our VLBA 5-GHz observations detected two compact core structures and confirmed the strong contributions from jets on their parsec-scale radio emission.
The jets of the other undetected sources lack compactness on parsec scales.

A magnetically heated corona is thought to power the compact nuclear radio emission in radio-quiet AGN. 
A diagnostic for coronal activity is the ratio of radio to X-ray luminosity, which satisfies the G\"{u}del–Benz relation, ($L_{R}/L_{X}$$\sim$10$^{-5}$, \citealt{Gudel1993}), and exhibiting a flat spectrum on high radio frequency \citep{laor2008origin,panessa2019origin}. 
The Chandra X-ray observations of the four dual AGN argue against a disk-corona model for the origin of nuclear X-rays due to their low Eddington ratios ($-$5$<$$\mathrm{log}$ $\lambda_\mathrm{X}$$<$$-$3, \citealt{2019ApJ...883...50G}). 
The steep spectra of target sources also argue against the coronal origins of radio emission.
According to G\"{u}del–Benz relation, only J2206+0003A have about 6\% radio emission originating from the corona, the contribution of the corona to the radio emission in the other sources less than 1\% and can be ignored.

For a wind origin, shocks caused by the outflow interactions with the interstellar medium are expected to ionize and expel the gas to produce blue-shifted X-ray, optical, and molecular lines \citep{panessa2019origin}. 
The Keck/LRIS optical spectral analysis indicates that all spectra with measurable emission lines can be well fit with tied emission lines \citep{gross2023testing}. 
This means that there is no direct evidence of the wind/outflow origin. But we also can't rule out it as a possible origin of extended radio emission.
Wide-angle winds/outflows would appear with many clumpy features in VLBI images \citep[e.g.,][]{yang2021nearby}. 
However, no apparent clumpy structure was found in our VLBA observations, which may be due to the non-detection of the weak radio emission of our targets.

Star formation produces extended radio emission on galactic scales, most of which would be resolved in high-resolution observations.
After dust extinction and aperture loss correction, the H$\alpha$ luminosity should be a good tracer of the total star formation rate (SFR).
Therefore, we estimated the SFRs of targets using their dust-extinction and aperture-loss corrected H$\alpha$ luminosities based on the formula \citep{Murphy2011}:
\begin{equation}\label{formula3}
\mathrm{SFR}_\mathrm{H\alpha} \thinspace (M_{\odot} \thinspace \mathrm{yr^{-1}}) = \frac{L_{H\alpha}}{5.37 \times 10^{42} \thinspace \mathrm{erg/s}}
\end{equation}

It is noted that the H$\alpha$-based SFRs should be upper-limit values because of the contribution on H$\alpha$ luminosity of AGN-photoionized gas.
The 1.4-GHz radio luminosity can be used as an SFR estimator, which is justified by the tight radio–far infrared (FIR) correlation \citep[e.g.,][]{Yun2001}.
Therefore, we can predict the expected 1.4-GHz radio power from star formation alone based on their SFRs \citep{Hopkins2003}:
\begin{equation}\label{formula4}
\mathrm{SFR}_\mathrm{1.4GHz} \thinspace (M_{\odot} \thinspace \mathrm{yr^{-1}}) = \frac{fL_{1.4GHz}}{1.81 \times 10^{21} \thinspace \mathrm{W/Hz}}
\end{equation}

where
\begin{equation}\label{formula5}
f = \begin{cases}
1 &  (L_\mathrm{1.4GHz} > L_c) \\
\left[0.1 + 0.9 \left(\frac{L_\mathrm{1.4GHz}}{L_c}\right)^{0.3}\right]^{-1} &  (L_\mathrm{1.4GHz} < L_c)
\end{cases}
\end{equation}

and $L_c$ = 6.4×10$^{21}$ W/Hz.
Then, we assumed the spectral index $\alpha$=$-$0.8 to infer the non-thermal radio luminosity originating from star formation to the 5-GHz frequency \citep[e.g.,][]{Condon1992}.
The 5-GHz radio emission origins and their proportions of all targets are shown in Table \hyperlink{tab5}{5}. 
As can be seen in Table \hyperlink{tab5}{5}, the radio emission of almost all targets is jet-dominated, except J2206+0003B.

\hypertarget{tab5}{}
\begin{deluxetable*}{ccccc}[h]
\tablenum{5}
\tablecaption{The various origins of 5-GHz radio emission in the four dual AGN}
\tablewidth{0pt}
\tablehead{
\colhead{Source} & \colhead{Compact jet} & \colhead{Extended jet+wind} &
\colhead{Corona} & \colhead{Star formation} 
}
\startdata
J0051+0020A & $<$13\% & $>$64\%  & -- & $<$23\%  \\
J0051+0020B & $\sim$56\% & $>$25\% & --  & $<$19\% \\
J2206+0003A & $<$20\% & $>$47\%  & $\sim$6\% & $<$27\%  \\
J2206+0003B & ? & ?  & -- & $<$72\% \\
J2232+0012A & $<$51\% & $>$49\%  & -- & -- \\
J2232+0012B & $<$81\% & $>$19\%  & -- & -- \\
J2300$-$0005A  & $\sim$71\% &  $\sim$29\% & -- & -- \\
J2300$-$0005B & ? & ?  & -- & -- \\
\enddata
\tablecomments{The compact-jet proportion is from the VLBA 5-GHz detection. The contribution of the corona is estimated through G\"{u}del–Benz relation. The star-formation proportion is estimated based on the H$\alpha$ luminosity and formula (\ref{formula3})--(\ref{formula5}). The rest of the radio emission is thought to originate from extended jet and wind. The question mark represents the unknown proportion based on current observations. The short line represents the contribution is less than 3\% and should be ignored. The corrected H$\alpha$ and 5-GHz luminosity data used in the calculation is from \cite{2019ApJ...883...50G}.}
\end{deluxetable*}

The radio emission properties of J0051+0020B exhibit signatures typical of AGN jets: a compact parsec-scale structure and a high brightness temperature. 
The VLBA detected a mas-scale core comprising 56\% of the total radio emission. 
The compact radio core has a high brightness temperature consistent with non-thermal synchrotron emission from a jet. 
Moreover, J0051+0020B exhibits a significant star formation on kiloparsec scales given its radio steep spectrum and the status of AGN/starburst composite in the BPT [N II] classification. 
The contribution of star formation to radio emission is about 19\%.
The compact jet likely dominates the observed radio output, while star formation makes secondary contributions on larger scales.
The characteristics of J0051+0020B, particularly its high accretion rate and low black hole mass, resemble those of Narrow-Line Seyfert 1 (NLS1) galaxies, some of which have been observed to exhibit pronounced jet activity (e.g., I Zw 1 and Mrk 335, \citealt{wang2023vlbi2}). 
This appears to be consistent with a scenario in which galaxy merging has triggered the black hole accretion and jet activity. 

The radio emission of J2300$-$0005A shows a mas-scale radio core which accounts for approximately 71\% of its total radio emission in our VLBA 5-GHz observations.
The contribution of star formation and corona should be ignored. 
The remaining about 30\% of the radio emission mostly originates from the extended jet and wind. 
J2300$-$0005A has a relatively lower Eddington ratio (log$\lambda_\mathrm{[O\thinspace \uppercase\expandafter{\romannumeral3}]}$=--4.0±0.7) and a higher radio loudness, suggesting the black hole is not actively consuming a significant amount of material.

Although the other six sources were undetected by VLBA, their compact structures in the VLA C-band observations support the existence of kiloparsec-scale jet originating from AGN. 
However, the non-detection of J2206+0003B and J2300$-$0005B should be due to insufficient sensitivity of the VLBA, so we cannot determine their radio emission compactness.
Among them, the contribution of star formation in J2206+0003B should be significant at 5 GHz (see Table \hyperlink{tab5}{5}).
In the other four VLBA-undetected sources, although star formation and corona have some contribution to total radio emission, they are still jet-dominated and similar to Seyfert galaxies or low luminosity AGN (LLAGN).
It is noted that the two components of J2232+0012 lack detection in the Gaia, VLBA and Chandra X-ray observations, with the largest separation ($\sim$11.6 kpc, see Figure \hyperlink{fig2}{2}) among the four dual AGN systems. 
They might represent an earlier stage of galaxy merging in which the AGN accretion may not have been significantly enhanced yet. 
As the galaxies continue to merge and interact, the accretion activity of the AGN could potentially increase, leading to more significant emission.

The diverse properties suggest these dual AGN may span a range of evolutionary stages, with star formation or AGN radio activity selectively enhanced at different merger phases \citep[e.g.,][]{hopkins2006fueling}. 
The AGN accretion could be the result of the combination of merger-driven accretion and stochastic accretion \citep[e.g.,][]{2019ApJ...883...50G, Steffen2023ApJ}.
The multi-wavelength data provides clues to unravel the complex interplay between gas dynamics, star formation, and black hole growth in these galaxy interactions. 
More observations are needed to draw more definitive, quantitative conclusions. Nonetheless, these first VLBA detections demonstrate the power of high-resolution VLBI imaging to detect compact radio cores and locate the positions of supermassive black holes.

\subsection{Multi-band and Multi-scale Analysis} \label{discussion3}

\begin{figure}[ht!]
\centering
\includegraphics[height=8.8cm, width=17.5cm]{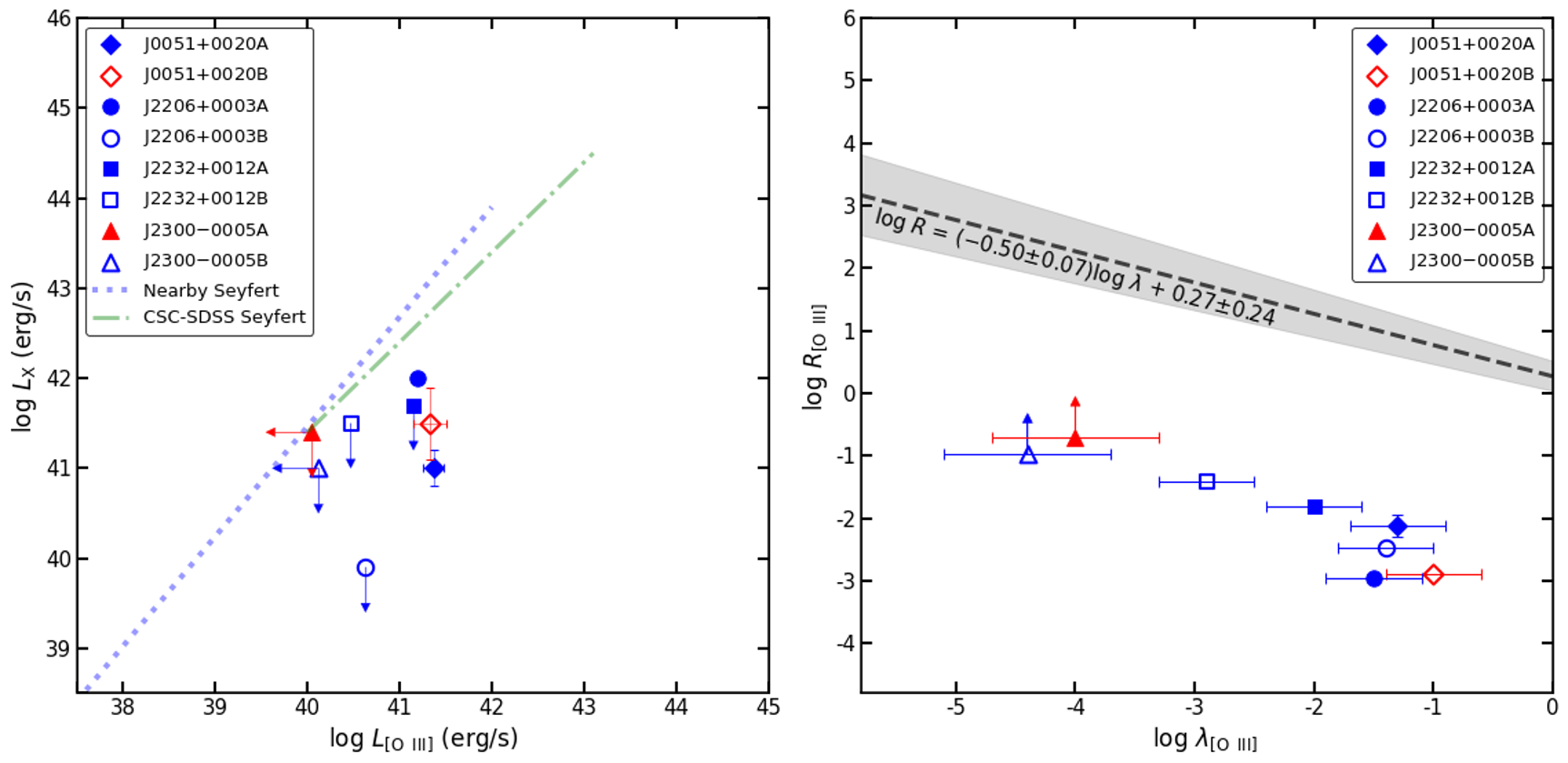}
\caption{
The multi-band emission properties of the target sources. Left panel: unabsorbed rest-frame 2--10 keV X-ray luminosity ($L_\mathrm{X}$) vs. corrected [O\thinspace \uppercase\expandafter{\romannumeral3}]\thinspace $\lambda$5007 emission-line luminosity ($L_\mathrm{[O\thinspace III]}$). The green dash-dot line represents the relation of Type 2 Seyfert galaxies from the CSC-SDSS cross-match catalogue \citep{Trichas2012}. The blue dot line is the relation for nearby Seyfert galaxies from \cite{Panessa2006}. Right panel: radio loudness ($R_\mathrm{[O\thinspace \uppercase\expandafter{\romannumeral3}]}$) vs. Eddington ratio ($\lambda_\mathrm{[O\thinspace \uppercase\expandafter{\romannumeral3}]}$), where $R_\mathrm{[O\thinspace \uppercase\expandafter{\romannumeral3}]}$ is the VLA 1.4-GHz to [O\thinspace \uppercase\expandafter{\romannumeral3}] luminosity ratio, $\lambda_\mathrm{[O\thinspace \uppercase\expandafter{\romannumeral3}]}$ is calculated $\lambda_\mathrm{[O\thinspace \uppercase\expandafter{\romannumeral3}]}$ $ =(L_\mathrm{mech}+L_\mathrm{rad})/L_\mathrm{Edd}$, $L_\mathrm{mech} = 10^{43}$ $ \mathrm{erg}$ $ \mathrm{s^{-1}} (L_{1.4 \mathrm{GHz}}/10^{24}\mathrm{W} \mathrm{Hz^{-1}})^{0.7}$ represents the jet mechanical luminosity \citep{Cavagnolo2010}, $L_\mathrm{rad} = 3500L_\mathrm{[O\thinspace \uppercase\expandafter{\romannumeral3}]}$ represents the radiative luminosity \citep{Heckman2004}. 
The black dashed line shows the inversely correlated relation between radio loudness and Eddington ratio, and the grey shadow represents its error range \citep{2002Ho}. The X-ray data and [O\thinspace \uppercase\expandafter{\romannumeral3}] data are from \cite{2019ApJ...883...50G}. The VLA 1.4-GHz data is from \cite{fu2015radio}. \label{fig:4}}
\end{figure}

Multi-band datasets of the four dual AGN observed by the SDSS, VLA, Chandra and Keck allow us to study their multi-band emission properties.
According to the Chandra X-ray observations by \cite{2019ApJ...883...50G}, three of the eight components, J0051+0020A\&B and J2206+0003A, have definite AGN-dominated X-ray emission.
Three sources (J2206+0003B, J2300-0005A\&B) have significant non-nuclear X-ray activity, such as hot interstellar medium (ISM) or X-ray binaries (XRBs).
J2232+0012A\&B are not detected by Chandra.
For the latter five sources, their X-ray luminosities should be considered as upper limits of the AGN X-ray luminosity.
The unabsorbed rest-frame 2--10 keV luminosities and [O\thinspace III] luminosities corrected for reddening and aperture loss are shown in the left panel of Figure \ref{fig:4}.
All of the target sources seem to have lower X-ray-to-[O\thinspace \uppercase\expandafter{\romannumeral3}] luminosity ratios, compared to typical isolated Seyfert galaxies (blue dot line and green dash-dot line, \citealt{Panessa2006, Trichas2012}).
\cite{Liuxin2013} found a similar result in four double-peaked [O\thinspace III]-selected kiloparsec-scale dual AGN.
They think the systematic X-ray deficit may be caused by a combination of higher nuclear gas column and viewing angle bias.
In the merger systems, the gas inflow driven by the tidal torques could increase the nuclear gas column and result in the X-ray deficit \citep[e.g.,][]{Hernquist1989}.
Of the three AGN-dominated X-ray sources, the VLBA-detected source J0051+0020B has high compactness ($S_\mathrm{VLBA}/S_\mathrm{VLA}>$50\%, see Table \hyperlink{tab5}{5}) indicating it should be principally face-on.
Therefore, the X-ray deficit of J0051+0020B could be caused by merger-driven gas inflow.
The other two AGN-dominated X-ray sources were undetected by VLBA and could have a large viewing angle.
Therefore, in addition to a possible gas inflow, their X-ray deficit should be caused by the viewing angle effect.

\cite{2019ApJ...883...50G} think the $L_\mathrm{[O\thinspace III]}$ enhancement from star formation could also cause low X-ray-to-[O\thinspace III] luminosity ratios.
On about 10-kiloparsec scales, the merger-driven gas inflow could promote star formation in the center of galaxies \citep[e.g.,][]{Patton2013, Ellison2013}.
The merger-driven star-formation enhancement could be influenced by many factors, such as stellar mass, projected separation, and relative inclination.
The SDSS and Keck/LRIS optical spectra of three dual AGN in targets exhibit $\sim$40-Myr young star formation, which is consistent with they are mostly AGN--star-forming composite \citep{fu2015radio}.
The other pair (J2300$-$0005A\&B) has an optical spectrum consistent with purely old stellar populations ($\sim$10 Gyr).
To study the contribution of star formation to the [O\thinspace III] luminosity, we plotted the relation between radio loudness and Eddington ratio based on the $L_\mathrm{1.4 GHz}$ and $L_\mathrm{[O\thinspace III]}$ in the right panel of Figure \ref{fig:4}.
The target sources approximatively exhibit the trend of the increase of radio loudness with decreasing Eddington ratios similar to AGN \citep[e.g.,][]{2002Ho, 2005Nagar}.
Therefore, in our case, the [O\thinspace III] luminosities of most targets are still AGN-dominated instead of star-forming dominated.

\begin{figure}[ht!]
\centering
\includegraphics[height=8.52cm, width=9cm]{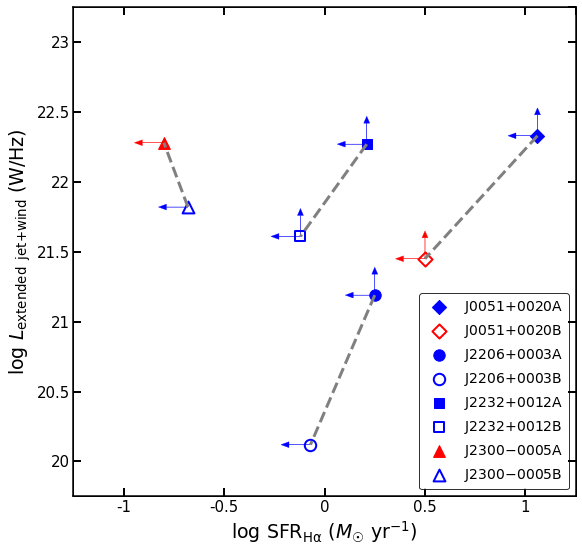}
\caption{
The 5-GHz radio emission from extended jet and wind vs. star formation rate. The radio luminosity of extended jet and wind is estimated according to Table \hyperlink{tab5}{5}. For J2206+0003B and J2300$-$0005A, we assume their all radio emission other than star formation originate from extended jets or wind. \label{fig:5}}
\end{figure}

To study the feedback mode of dual AGN, we examined the relation between extended AGN radio emission and star formation rates in host galaxies, shown in Figure \ref{fig:5}.
For the three AGN pairs with young star formation, the star formation rates exhibit positive correlations with their radio luminosities originating from extended jet and wind in each pair.
It seems to imply positive feedback between dual AGN and their host galaxies.
The other pair J2300$-$0005 almost has no star formation and is very different from the other three pairs.
They could be in different stages of merging and exhibit different properties of the black holes and host galaxies.
A more detailed study requires the observations and studies of larger samples in the future.

In the multi-band and multi-scale analysis, the emission properties of dual AGN could exhibit different (e.g., Eddington ratio, radio compactness) or similar (e.g., AGN activation, X-ray deficit) with isolated AGN.
The merger does seem to facilitate low to moderate levels of AGN accretion. 
Still, meanwhile, the AGN in the merger systems may also exhibit large differences \citep[e.g.,][]{Van2012}.
Overall, the multi-band emission properties of target sources do not seem to be explained by just one model of galactic mergers.
Tidally induced effects and stochastic triggers could work simultaneously during the merger process \citep[e.g.,][]{Steffen2023ApJ}.

\newpage
\section{Conclusions} \label{Conclusions}

We observed four dual-AGN systems using the VLBA at 5 GHz to obtain their high-resolution images. 
The primary goal was to investigate their radio emission properties on parsec scales to understand the possible origins of the emission in dual AGN. 
Our conclusions are summarized below:

\begin{enumerate}
\item[1).] Two radio cores, labeled as J0051+0020B and J2300$-$0005A, were clearly detected on parsec scales for the first time in the two dual-AGN systems. 
We provided their precise coordinates, flux densities, brightness temperatures and radio emission powers. 
Combining the results of multi-band observations, we conclude that the radio emission in J0051+0020B and J2300$-$0005A is primarily dominated by compact jets.

\item[2).] For the other six undetected AGN, most of the radio emission was resolved out in our high-resolution images. 
We provided the upper limits of the flux densities, brightness temperatures, and emission powers for these extended sources. 
We deduce that their radio emission could arise from extended jets or outflows.
For J2206+0003B, star formation also plays a role in its 5-GHz radio emission.

\item[3).] Our kiloparsec-scale dual-AGN sample exhibits a systematically lower X-ray-to-[O\thinspace III] luminosity ratio compared to isolated Seyfert galaxies. Our VLBA observations suggest the X-ray deficit is likely attributed to the tidally induced effect and possible viewing angle effect. 

\end{enumerate}

\acknowledgments
We thank Hai Fu and Masoumeh Ghasemi Nodehi for helpful discussions and useful suggestions. This work is supported by the CAS `Light of West China' Program (grant No. 2021-XBQNXZ-005), the National SKA Program of China (grant No. 2022SKA0120102) and the National Key R\&D Program (grant Nos. 2018YFA0404602 and 2023YFE0102300). L.C. acknowledges the support from the Tianshan Talent Training Program (grant No. 2023TSYCCX0099). T.A., L.C.H., and N.C. acknowledge the support from the Xinjiang Tianchi Talent Program. T.A. thanks to the open funding of the Pinghuy Laboratory. This work is also partly supported by the Urumqi Nanshan Astronomy and Deep Space Exploration Observation and Research Station of Xinjiang (XJYWZ2303). L.C.H. was supported by the National Science Foundation of China (11721303, 11991052, 12011540375, 12233001), the National Key R\&D Program of China (2022YFF0503401), and the China Manned Space Project (CMS-CSST-2021-A04, CMS-CSST-2021-A06). The VLBA is a joint facility of independent North American radio astronomy institutes. Scientific results from data presented in this publication are derived from the following VLBA project codes: BL255A \& BL255B 
(P.I.: X. Liu). 
This work has made use of data from the European Space Agency (ESA) mission {\it Gaia} (\url{https://www.cosmos.esa.int/gaia}), processed by the {\it Gaia} Data Processing and Analysis Consortium (DPAC, \url{https://www.cosmos.esa.int/web/gaia/dpac/consortium}). Funding for the DPAC has been provided by national institutions, in particular the institutions participating in the {\it Gaia} Multilateral Agreement.

\bibliography{arxiv}{}
\bibliographystyle{aasjournal}

\end{document}